\documentclass{JHEP3}
\title{Noncommutative Quantum Field Theory:\\A
Confrontation of Symmetries}

\author{M. Chaichian$^a$, K. Nishijima$^b$, T. Salminen$^a$ and A. Tureanu$^a$\\
$ ^a$Department of Physics, University of Helsinki and Helsinki
Institute of Physics, \\P.O. Box
64, 00014 Helsinki, Finland\\

$^b$ Department of Physics, University of Tokyo 7-3-1 Hongo,\\
Bunkyo-ku, Tokyo 113-0033, Japan}

\abstract{The concept of a noncommutative field is formulated based
on the interplay between twisted Poincar\'e symmetry and residual
symmetry of the Lorentz group. Various general dynamical results
supporting this construction, such as the light-wedge causality
condition and the integrability condition for Tomonaga-Schwinger
equation, are presented. Based on this analysis, the claim of the
identity between commutative QFT and noncommutative QFT with twisted
Poincar\'e symmetry is refuted.}

\usepackage{amsmath,amssymb,multirow,latexsym,amsfonts}
\newcommand{\be}{\begin{equation}}
\newcommand{\ee}{\end{equation}}
\newcommand{\bea}{\begin{eqnarray}}
\newcommand{\eea}{\end{eqnarray}}

\newcommand {\vect}{\mathbf}
\newcommand{\R}{{\mathbb R}}

\def\P{Poincar\'e }
\begin{document}
\renewcommand {\theequation}{\thesection.\arabic{equation}}
\renewcommand {\thefootnote}{\fnsymbol{footnote}}

\section{Introduction}
\setcounter{equation}{0}

Symmetry principles are invaluable guiding tools in the formulation
of physical theories. Lie-group based symmetries proved their full
worth in the construction of relativistic quantum field theory, in
gauge field theories - actually in all the experimentally proven
theories that are known. The more exotic quantum groups have been
exhaustively studied starting with the 80ties; they have been mostly
explored in various deformations of quantum mechanical systems, but
not in the formulation of field theories. It is therefore natural
that one particular deformation used in quantum groups, the twist,
has become very popular since the twisted \P algebra was put in
connection with the actively studied noncommutative field theories
\cite{CKNT}. The connection between noncommutative space-time
\cite{DFR,SW} and quantum symmetry has its precursors in the context
of string theory and quasitriangular Hopf algebras \cite{Watts},
followed shortly by an approach in the dual language of Hopf
algebras \cite{oeckl}. What especially attracted interest when the
twisted \P algebra was rediscovered as a symmetry of noncommutative
space-time was the realization that its representation content is
the same as the one of the usual \P algebra \cite{CKNT}. At the time
there was a well known problem conserning the representation content
of the theory, which was now solved by the discovery of this new
symmetry. The problem, as perceived earlier (see, for example,
\cite{LAG_ax}), was that NC QFT on four-dimensional space-time was
known to be symmetric under a subgroup of the Lorentz group, $SO
(1,1)\times SO(2)$ (in case the time is noncommutative) or
$O(1,1)\times SO(2)$ (which contains also reflection and is valid in
case time is commutative) \cite{LAG}, which are both Abelian groups
and thus have only one-dimensional irreducible representations.
Thus, the notion of spin seemed to be irremediably lost in NC QFT.
Twisted \P algebra rescued the spin of the representations and,
moreover, indicated that one-particle irreducible representations
retain the same classification in terms of mass and spin as in the
case of \P symmetry. Twisted \P symmetry became thus a new concept
of relativistic invariance for NC QFT \cite{CPrT}.

Another thing that made the twist deformation very alluring was its
simplicity. For the consistency of argumentation proposed in this
paper we shall repeat a few main formulas of the construction of the
twisted \P algebra (for details on twist deformations and other
quantum group techniques, see Refs. \cite{ChPr,Maj,ChD}). The
twisted \P algebra is the universal enveloping of the \P algebra
$\cal U(\cal P)$, viewed as a Hopf algebra, deformed with the
Abelian twist element \cite{drinfeld}
\be\label{abelian twist}{\cal
F}=\exp\left({\frac{i}{2}\theta^{\mu\nu}P_\mu\otimes
P_\nu}\right),\ee
where $\theta_{\mu\nu}$ is a constant antisymmetric matrix (and not
a tensor, i.e. it does not transform under the Lorentz
transformations) and $P_\mu$ are the translation generators. This
induces on the algebra of representations of the \P algebra the
deformed multiplication,
\be\label{twist prod} m\circ(\phi\otimes\psi)=\phi\psi\rightarrow
m_\star\circ(\phi\otimes\psi)=m\circ{\cal
F}^{-1}(\phi\otimes\psi)\equiv \phi\star\psi\,,\ee
which is precisely the well known Weyl-Moyal $\star$-product (taking the
Minkowski space realization of $P_\mu$, i.e.
$P_\mu=-i\partial_\mu$):
\be\label{star}\star=\exp{\left(\frac{i}{2}\theta^{\mu\nu}\overleftarrow\partial_\mu\overrightarrow\partial_\nu\right)}\
.\ee

The twist (\ref{abelian twist}) does not affect the actual
commutation relations of the generators of the \P algebra $\cal P$:
\bea [P_\mu,P_\nu]&=&0,\cr
[M_{\mu\nu},P_\alpha]&=&-i(\eta_{\mu\alpha}
P_\nu-\eta_{\nu\alpha}P_\mu),\cr
[M_
{\mu\nu},M_{\alpha\beta}]&=&-i(\eta_{\mu\alpha}M_{\nu\beta}-\eta_{\mu\beta}M_{\nu\alpha}-\eta_{\nu\alpha}M_{\mu\beta}+\eta_{\nu\beta}M_{\mu\alpha}).
 \eea
Consequently also the Casimir operators remain the same and the
representations and classifications of particle states are identical
to those of the untwisted \P algebra.

However, the twist deforms the action of the generators in the
tensor product of representations, or the so-called {\it coproduct}.
In the case of the usual \P algebra, the coproduct $\Delta_0\in\cal
U(\cal P)\times \cal U(\cal P)$ is symmetric,
\be\label{primitive} \Delta_0(Y)=Y\otimes1+1\otimes Y,\ee
for all the generators $Y\in \cal P$. The twist $\cal F$ deforms the
coproduct $\Delta_0$ to $\Delta_t\in {\cal U}_t({\cal P})\times
{\cal U}_t(\cal P)$ as:
\be\label{twist_coproduct}\Delta_0(Y)\longmapsto\Delta_t(Y)={\cal
F}\Delta_0(Y){\cal F}^{-1}\,.\ee
This similarity transformation is compatible with all the properties
of $\cal U(\cal P)$ as a Hopf algebra, since $\cal F$ satisfies the
twist equation:
\be\label{twist_eq} {\cal F}_{12}(\Delta_0\otimes id){\cal F}={\cal
F}_{23}(id\otimes\Delta_0){\cal F}\,, \ee
where ${\cal F}_{12}={\cal F}\otimes 1$ and ${\cal F}_{23}=1\otimes
{\cal F}$.

The twisted coproducts of the generators of \P algebra turn out to
be:
\bea \Delta_t(P_{\mu})&=& \Delta_0(P_{\mu})=P_{\mu}\otimes
1+1\otimes P_{\mu},\label{twist p}\\
\Delta_t(M_{\mu\nu})&=&M_{\mu\nu}\otimes 1+1\otimes
M_{\mu\nu}\label{twist
m}\\
&- &\frac{1}{2}\theta^{\alpha\beta}
\left[(\eta_{\alpha\mu}P_\nu-\eta_{\alpha\nu}P_\mu)\otimes
P_\beta+P_\alpha\otimes
(\eta_{\beta\mu}P_\nu-\eta_{\beta\nu}P_\mu)\right].\nonumber\eea
Thus the twisted coproduct of the momentum generators is identical
to the primitive coproduct, eq. (\ref{twist p}), meaning that
translational invariance is preserved, while the twisted coproduct
of the Lorentz algebra generators, eq. (\ref{twist m}), is
nontrivial, implying the violation of Lorentz symmetry.

Taking in (\ref{twist prod}) $\phi(x)=x_\mu$ and $\psi(x)=x_\nu$,
one obtains:
\be [x_\mu, x_\nu]_\star=i\theta_{\mu\nu}. \ee
This is the usual commutation relation of the Weyl symbols of the
noncommuting coordinate operators $\hat x$,
\be [\hat x_\mu, \hat x_\nu]=i\theta_{\mu\nu}, \ee
which is obtained in the Weyl-Moyal correspondence. Thus, the
construction of a NC quantum field theory through the Weyl-Moyal
correspondence is equivalent to the procedure of redefining the
multiplication of functions, so that it is consistent with the
twisted coproduct of the \P generators (\ref{twist_coproduct})
\cite{CKNT}.

As such, one would expect that all the features obtained in the past
for NC QFT, like the connection between topology of space-time and
the UV behaviour \cite{CDP}, UV/IR mixing \cite{UVIR}, the light-wedge
causality condition \cite{LAG,CNT,Chu}, preservation of CPT symmetry
and spin-statistics relation \cite{CNT}, formulation of
noncommutative gauge theories with symmetry under $\star$-gauge
transformations \cite{Haya} obeying very strict rules \cite{nogo},
Lorentz-symmetry violation of the $S$-matrix in interacting NC
theory \cite{q-shift} etc. would be confirmed by the symmetry of NC
QFT under twisted \P algebra.

The alluring simplicity of the twist turned it into the key-concept
based on which noncommutative (quantum) field theories and
noncommutative gravity have lately been studied. The recipe for
extending the twist to other symmetries, like gauge symmetries and
diffeomorphism transformation, seemed also at hand: one had to
consider a commutative model with a certain symmetry, extend that
symmetry by the \P algebra through a direct or semidirect product,
and use the twist element (\ref{abelian twist}) to deform the new
enveloping algebra. The $\star$-product (\ref{star}) would
automatically appear instead of the usual multiplication, due to
(\ref{twist prod}), and the result would be a noncommutative gauge
theory, for instance, with twisted gauge symmetry.

Using the twist deformation by the Abelian twist  (\ref{abelian
twist}) and also the prescription given above, new results appeared
in the literature, contradicting all the above mentioned features of
noncommutative quantum field and gauge theories: the UV/IR mixing
allegedly disappeared \cite{bal-UVIR}; the spin-statistics relation
was claimed not to hold \cite{bal-spin}; twisted diffeomorphisms
seemed to provide the general coordinate transformations in
noncommutative gravity constructed with an immutable
coordinate-dependent $\star$-product \cite{Wess-grav};
noncommutative gauge theories appeared to be easily constructed with
symmetry under any gauge group (not only $U_\star(n)$) and
possessing any representations \cite{Vassilevich,Wess};  until
finally NC QFT seemed to have the usual, light-cone causality
condition as well as Lorentz symmetry, and ultimately to be
identical to commutative QFT \cite{FW,Abe,Aschieri}.

The various controversies that ensued were resolved in favour of the
traditional dynamical approach to NC QFT: the UV/IR mixing was shown
still to be present \cite{AT,Zahn}; the spin-statistics relation was
proven to hold \cite{AT,yee,AT-kyoto}; the twisted gauge theories
and implicitly the twisted diffeomorphisms were shown to be
constructed in a manner inconsistent with the concept of gauge
invariance \cite{CT,CTZ}, thus leaving only the option of
$\star$-gauge symmetry with its restrictions. The consistent use of
the twist deformation technique turned out to support the dynamical
calculations. We shall not return to these issues in this paper.

In this paper we shall show the intrinsic impossibility of the
identity between noncommutative and commutative (quantum) field
theory. We shall approach the subject from different points of view:
a general argument, based on Pauli's Theorem; a general derivation
of the causality condition in noncommutative interacting theories as
integrability condition for the Tomonaga-Schwinger equation; and
finally a new interpretation of the noncommutative field operator
itself in a theory with twisted \P symmetry. All these approaches
will lead to the same conclusion, that the twisted \P symmetry of
noncommutative (quantum) field theory is reduced to the residual
$O(1,1)\times SO(2)$ symmetry, but still carrying representations of
the full Lorentz group. Consequently, Lorentz invariance is absent
and noncommutative QFT is in essence different from commutative QFT.

\section{Lorentz invariance and Pauli's Theorem}
\setcounter{equation}{0}

In 1957, after learning that weak interactions violate parity, Pauli
introduced what we shall call the Pauli group (not to be confused
with the group of the $\sigma$-matrices!)  in order to explain why
the violation of parity had not been earlier recognized in
beta-decay \cite{Pauli}. In our case we shall use not the Pauli
group itself, but the philosophy behind it, as described in
\cite{KN-pauli}.

Let ${\cal L}(g_i;\psi_j)$ be the Lagrangian density of a system of
fields, where $\psi_j$ denotes the field operators and $g_i$ a
fundamental parameter, such as a mass, or a coupling constant, or -
in our case - the noncommutativity parameter. Assuming that the
field operators transform under a group $G$ as
\be\label{field-transform}\psi_j\to \psi'_j,\ee
the change in ${\cal L}$ caused by (\ref{field-transform}) can be
compensated by a change in the parameters,
\be\label{g-transform}g_i\to g'_i,\ee
such that the Lagrangian density would be invariant,
\be\label{L-inv}{\cal L}(g_i;\psi_j)={\cal L}(g'_i;\psi'_j).\ee

An observable quantity will depend on the set of parameters in the
Lagrangian, but not on the field operators. If $G$ is a symmetry group
of the system described by the Lagrangian ${\cal
L}(g_i;\psi_j)$, then an observable ${\cal O}(g_i)$ must satisfy the
condition:
\be\label{obs-inv}{\cal O}(g_i)={\cal O}(g'_i),\ee
in other words, ${\cal O}(g_i)$ must be a function of $g_i$
invariant under $G$.

The statement \eqref{obs-inv} will be called Pauli's Theorem in what
follows. There is a question whether this theorem is valid for the
Lorentz group or not. Practical calculations show that the
$S$-matrix elements depend not only on Lorentz invariant
combinations such as $\theta^{\mu\nu}\theta_{\mu\nu}$, but also on
non-invariant $p^\mu\theta_{\mu\nu} k^\nu$,
$p^\mu\theta_{\mu\alpha}\theta^{\alpha\nu} k_\nu$ etc., indicating
violation of Pauli's Theorem. It is plausible, therefore, that
Pauli's Theorem is valid only for internal symmetry group $G$ and
for a finite set of parameters $\{g_i\}$, just like the
Coleman-Mandula Theorem. The momenta $p$ ,$k$,... are not present in
the original Lagrangian and they can not be included in the finite
set of parameters, showing explicitly violation of Pauli's Theorem
for the Lorentz group\footnote{If one erroneously applies this
theorem to the Lorentz group one may come to the conclusion of
Lorentz symmetry for NC QFT. For example, in Ref. \cite{FW}, such a
conclusion was drawn in the axiomatic approach to NC QFT. While
justly observing that the shifts of coordinates $\star$-commute
among themselves and the noncommutative Wightman functions, as
translationally invariant objects, depend only on shifts of
coordinates, it was however overlooked that the shifts of
coordinates contracted with $\theta_{\mu\nu}$ are also commuting
variables which may (and will) appear in the Wightman functions.
Indeed, by shifting the coordinates in a $\star$-product of
functions, the $\theta$-dependence does not vanish. Should the
$\theta$-dependence of the Wightman functions disappear by the shift
of coordinates, it would mean that the requirement of translational
invariance implies necessarily Lorentz invariance.}.

Based on this general argument we have to conclude that Lorentz
invariance is violated in NC QFT with twisted \P symmetry, if the
parameter $\theta_{\mu\nu}$ appears in the observables. The complete
disappearance of $\theta_{\mu\nu}$ from the observables or its
presence contracted only to itself would be the effect of a peculiar
conspiracy of accidents.

In actual calculations performed in NC QFT the appearance of
$\theta_{\mu\nu}$ contracted with momenta of particles is
commonplace, and it is the reason for the emergence of UV/IR mixing
\cite{UVIR} and of the light-wedge causality condition
\cite{LAG,CNT,Chu}, to name only two essential aspects with
far-reaching consequences, among which the failure of analyticity of
the scattering amplitude \cite{LS} and the nonexistence of
high-energy bounds of the Froissart-Martin type on the total
cross-section in NC QFT \cite{Froissart bound} are representative
examples.

\section{The light-wedge causality condition and the Tomonaga-Schwinger equation in NC QFT}
\setcounter{equation}{0}

In anticipation of the light-wedge causality condition, we shall
consider that the constant matrix $\theta$ has no time-space
components, i.e. $\theta_{0i}=0$, compatible with causality
\cite{SST} and unitarity \cite{unit}. Without loss of generality, we
choose the coordinate system (which will be used throughout the
paper) in such a way that the $\theta$-matrix is written in the
form:
\begin{eqnarray}\label{theta}
\theta^{\mu\nu}=\left(
\begin{array}{cccc}
0 &0 & 0  & 0 \\
0 & 0 & 0  & 0 \\
0 & 0  &0 & \theta \\
0 & 0  & -\theta & 0
\end{array}
\right),
\end{eqnarray}
i.e. $\theta_{23}=-\theta_{32}=\theta$, while all other components
vanish. This configuration of the $\theta$-matrix is invariant under
the action of the subgroup $O(1,1)\times SO(2)$ of the Lorentz group
\cite{LAG}. When appropriate, we shall comment on other possible
configurations of the $\theta$-matrix as well. We further use the
notation
\begin{align}
&x^\mu = (\tilde x,\vect a), \,\,y^\mu = (\tilde y,\vect b)\,, \\
&\tilde x = (x^0,x^1), \, \tilde y = (y^0,y^1),\, \vect a =
(x^2,x^3), \, \vect b= (y^2,y^3)\,,
\end{align}
and consider $x^2,x^3$ (and $y^2,y^3$) as internal degrees of
freedom. We thus confine ourselves to one time and one space
dimension.

In the following we shall use the integral representation for the
Moyal $\star$-product, which reads, in general
\be\label{int rep}(f\star g)(x)=\int d^D y\ d^D z\ \mathcal K(x;y,z)
f(y) g(z)\,,\ee
where
\be \mathcal
K(x;y,z)=\frac{1}{\pi^{D}\det\theta}\exp[-2i(x\theta^{-1}y+y\theta^{-1}z+z\theta^{-1}x)]\,,\ee
with $D$ being the even dimension of the invertible matrix $\theta$,
$\det\theta$ being its determinant and we use the notation
$x\theta^{-1}y=x^\mu(\theta^{-1})_{\mu\nu}y^\nu$.

In our case, the invertible part of $\theta$ is a $2\times
2$-submatrix in the $(2,3)$-plane and the integration goes only over
the noncommutative coordinates, such that we can write the integral
form of the $\star$-product of $n$ functions as:
\be(f_1\star f_2\star\cdots\star f_n)(x)=\int d \vect a_1 d \vect
a_2\cdots d\vect a_n \mathcal K(\vect a;\vect a_1,\cdots,\vect a_n)
f_1(\tilde x,\vect a_1)f_2(\tilde x,\vect a_2)\cdots f_n(\tilde
x,\vect a_n)\,, \ee
where
\be\label{kernel} \mathcal K(\vect a;\vect a_1,\cdots,\vect
a_n)=\frac{1}{(\pi^{2}\det\theta)^{n/2}}\exp[-2i(\vect
a\theta^{-1}\vect a_1+\vect a_1\theta^{-1}\vect a_2+\cdots+\vect
a_n\theta^{-1}\vect a)]\,.\ee
The kernel \eqref{kernel} is $SO(2)$ invariant.

\subsection*{Tomonaga-Schwinger equation in two dimensions}

The Tomonaga-Schwinger equation \cite{tomonaga,schwinger} (see also
\cite{schweber}) is the covariant generalization of the Schr\"odinger
equation in the interaction picture, formulated as a functional
differential equation incorporating arbitrary Cauchy surfaces, and
not only those of constant Minkowski time.

In commutative QFT the Tomonaga-Schwinger equation reads:
\be\label{t-s} i\frac{\delta}{\delta \sigma(x)} \Psi[\sigma] =
\mathcal H_{int}(x)\Psi[\sigma]\,, \ee
where $\mathcal H_{int}(x)$ is the interaction Hamiltonian density,
and $\sigma$ is a space-like surface (i.e. a surface whose every two
points are space-like separated). The existence of solutions for the
Tomonaga-Schwinger equation is insured if the integrability
condition
\be\label{integr} \frac{\delta^2\Psi[\sigma]}{\delta
\sigma(x)\delta\sigma(x')}-\frac{\delta^2\Psi[\sigma]}{\delta
\sigma(x')\delta\sigma(x)}  = 0, \ee
with $x$ and $x'$ on the surface $\sigma$, is satisfied. This
integrability condition then implies
\be\label{inv com rule}[\mathcal H_{int}(x),\mathcal
H_{int}(x')]=0.\ee
Since in the interaction picture the field operators satisfy
free-field equations, they satisfy Lorentz invariant commutation
rules. The Lorentz invariant commutation relations are such that
(\ref{inv com rule}) is satisfied automatically, since $x$ and $x'$
are space-like separated.

In the noncommutative case, the use of the interaction picture has
the advantage that the free-field equations satisfied by the
noncommutative fields are identical to the corresponding free-field
equations of the commutative case. The Tomonaga-Schwinger equation
in the noncommutative case will read:
\begin{align}
&i\frac{\delta}{\delta \mathcal C} \Psi[\mathcal C] = \mathcal H_{int}(x)_\star\Psi[\mathcal C]\,, \\
&\mathcal H_{int}(x)_\star = \lambda[\phi(x)]_\star^n\,,
\end{align}
where $\mathcal C$ is a 1-dimensional surface (i.e. a curve)
embedded in the plane of commutative coordinates $(x^0,x^1)$. The
fields $\phi(x)$ satisfy free-field equations and the Hamiltonian of
interaction is built up by $\star$-multiplying the fields.

The integrability condition is:
\be \left[ \mathcal H_{int}(x)_\star,\mathcal H_{int}(y)_\star
\right] = 0\,, \quad \text{for} \,\, x,y\in \mathcal C\,, \ee
which we can write as
\begin{eqnarray}\label{NC integr cond}
\bigl[(\phi\star \ldots \star \phi)(\tilde x, \vect a), (\phi \star
\ldots \star \phi)(\tilde y, \vect b) \bigr]&=&\int \prod_{i=1}^n
d\vect a'_i \,\mathcal K(\vect a;\vect a'_1,\cdots,\vect a'_n) \int
\prod_{i=1}^n d\vect b'_i \, \mathcal K(\vect b;\vect
b'_1,\cdots,\vect b'_n) \cr
&\times&\bigl[\phi(\tilde x, \vect a'_1)\ldots \phi(\tilde x, \vect
a'_n), \phi(\tilde y, \vect b'_1)\ldots \phi(\tilde y, \vect b'_n)
\bigr]\cr
&=&0.
\end{eqnarray}
The commutators of products of fields appearing in (\ref{NC integr
cond}) are written as products of fields at various space-time
points multiplied by invariant commutators of fields. A typical
factor is
\be\phi(\tilde x, \vect a'_1)\ldots \phi(\tilde x, \vect
a'_{n-1})\phi(\tilde y, \vect b'_1)\ldots \phi(\tilde y, \vect
b'_{n-1})\bigl[\phi(\tilde x, \vect a'_n), \phi(\tilde y, \vect
b'_n) \bigr]\,.\ee
The fields at every point are independent, since they are systems
with an infinite number of degrees of freedom. As a result, their
products will also be independent. Eq. (\ref{NC integr cond})
becomes a sum of independent products of fields, whose coefficients
have to vanish identically in order for the whole sum to vanish.
Since the kernel can not vanish, it remains as a necessary condition
for the commutators of fields to be zero at every point,
\be \bigl[\phi(\tilde x, \vect a_i'), \phi(\tilde y, \vect b_j')
\bigr] =0 \,. \ee
This condition is satisfied outside of the mutual light-cone:
\be\label{light-cone cond} (x^0-y^0 )^2-(x^1-y^1 )^2 - ( {a_i^2}'-
{b_j^2}')- ({a_i^3}'-{b_j^3}' )^2 < 0 \,, \ee
since all $\phi(x)$ satisfy the same free-field equations and the
same invariant commutation relations as in the commutative case.
However, $\vect a_i'$ and $\vect b_j'$ are integration variables in
the range \be 0\leq ( {a_i^2}'-{b_j^2}')^2+( {a_i^3}'-{b_j^3}')^2 <
\infty \ee and therefore the necessary condition becomes
\be (x^0-y^0)^2-(x^1-y^1)^2 <0 \,, \ee
i.e. the light-wedge causality condition, symmetric under the
stability group of $\theta_{\mu\nu}$, $O(1,1)\times SO(2)$.

Remark that the light-wedge causality condition is obtained here in
a general approach, without using the actual mode expansion of the
fields, but only the fact that in the interaction picture the field
operators satisfy free-field equations and the integral
representation of the Moyal $\star$-product. In the Appendix we show
that the light-wedge configuration is obtained for any commutator of
$\star$-products of field operators, starting from the commutator
with the simplest powers of field operators.

We should point out that, were we to allow the time to be
noncommutative, i.e. $\theta_{0i}\neq 0$ (in a Lorentz invariant
manner), then time would have entered the $\star$-product and be
integrated over in the integral representation \eqref{int rep}. The
time variable as an integration variable which can not be fixed
would have crept in \eqref{light-cone cond}, resulting in the
impossibility of deriving any causality condition. We can therefore
conclude that quantum field theories with noncommutative time do not
fulfil an integrability condition for the Tomonaga-Schwinger
equation. Although some theories with noncommutative time may appear
to have desirable properties, like unitarity or Lorentz symmetry,
these constructions are jeopardized by the lack of solution of the
Tomonaga-Schwinger equation, implying that the space of states in
the interaction picture is empty.

The lack of causality is a problem also in certain theories in which
$\theta_{\mu\nu}$ is a Lorentz tensor and the Moyal $\star$-product
\eqref{star} is used \cite{Sami}. The hope of having wedge-causality
in a theory with $\theta_{\mu\nu}$ transforming as a Lorentz tensor
\cite{Lechner} may be deceptive: the shape of the wedge is given by
the commuting coordinates, since the nonlocality in the
noncommutative directions makes the speed of propagation of a signal
infinite in those directions. For a wedge to exist it is essential
that the time coordinate be commutative. Assuming that one starts
with a system of reference in which $\theta_{\mu\nu}$ has a form
similar to \eqref{theta} and wedge-locality is apparent, since
$\theta_{\mu\nu}$ is a genuine Lorentz tensor, one can always boost
to a frame of reference in which $\theta_{\mu\nu}$ picks up time
components. In the new system, time is noncommutative, consequently
the wedge simply disappears. In order to transform a wedge into
another by a Lorentz transformation, one has to discard all the
transformations which give $\theta_{\mu\nu}$ nonvanishing time
components, but this is to break the Lorentz symmetry from the
beginning.

The light-wedge causality condition for NC QFT has been recently
obtained in another general context, which is the axiomatic
formulation. Without any reference to specific models, based only on
the fact that the Wightman functions have to be defined in NC QFT
with $\star$-products \cite{axiomatic}:
$$
W_\star(x_1,x_2,...x_n)=\langle
0|\phi_1(x_1)\star\phi_2(x_2)\star...\star\phi_n(x_n)|0\rangle\,,
$$
it has been shown in \cite{tfCMTV} (see also \cite{soloviev}) that
the space of test functions smearing these noncommutative Wightman
functions is one of the Gel'fand-Shilov spaces ${S}^\beta$ with
$\beta<1/2$. These test functions can have finite support only in
the commutative directions (if such directions exist), therefore the
local commutativity condition (or microcausality condition) which is
central to the axiomatic approach can be formulated only with
respect to the light-wedge.

\section{Twisted \P symmetry and the residual $O(1,1)\times SO(2)$ invariance}
\setcounter{equation}{0}

The dynamical calculations performed or reviewed in this paper show
that noncommutative quantum field theories with a constant
noncommutativity parameter $\theta_{\mu\nu}$ break Lorentz
invariance and, depending on the structure of the $\theta$-matrix,
retain a symmetry under the Lorentz subgroup $SO(1,1)\times SO(2)$
when $\theta_{0i}\neq 0$ or under $O(1,1)\times SO(2)$ when
$\theta_{0i}=0$. The second case is of physical interest, since it
avoids the known problems with causality \cite{SST,LAG,CNT} and
unitarity \cite{unit}, preserving the notion of light-wedge
causality. General arguments, based on the philosophy of the Pauli
group, also support these results. It is then natural to expect that
a consistent construction based on the twisted \P algebra leads to
the same outcome\footnote{"Symmetry is a tool that should be used to
determine the underlying dynamics, which must in turn explain the
success (or failure) of the symmetry arguments. Group theory is a
useful technique, but it is no substitute for physics."(Howard
Georgi, \cite{georgi})}. A rigorous construction of noncommutative
fields, starting from first principles and twisted \P algebra, has
only recently been put forward \cite{CKTZZ}. Obviously, the
implications of twisted \P symmetry on the content of one-particle
irreducible representations should have bearing also on the
definition of noncommutative fields, though this relation is not
straightforward. Answering the question about what a noncommutative
field is, in the sense of the actions of the twisted \P algebra,
will finally lead us to the explicit meaning of twisted \P
invariance in NC QFT.

\subsection{$O(1,1)\times SO(2)$ invariance from the perspective of the twist}

Before moving further to the construction of a noncommutative
fields, let us first consider simple facts relating the twisted \P
algebra ${\cal U}_t({\cal P})$ and the residual symmetry
$O(1,1)\times SO(2)$. With the $\theta$-matrix configuration
\eqref{theta}, i.e. $x^0$ and $x^1$ as commutative coordinates and
$x^2$ and $x^3$ as noncommutative coordinates, one can calculate the
twisted coproducts of all the Lorentz generators according to the
formula \eqref{twist m}. The result is:
\bea\label{twist-resid} \Delta_t(M_{01})&=& \Delta_0(M_{01})=
M_{01}\otimes 1 + 1\otimes M_{01},\cr
\Delta_t(M_{23})&=& \Delta_0(M_{23})= M_{23}\otimes 1 + 1\otimes
M_{23}\,,\eea
while
\bea\label{twist-mix} \Delta_t(M_{02})&=&
\Delta_0(M_{02})+\frac{\theta}{2}(P_0\otimes P_3-P_3\otimes P_0),\cr
\Delta_t(M_{03})&=& \Delta_0(M_{03})-\frac{\theta}{2}(P_0\otimes
P_2-P_2\otimes P_0),\cr
\Delta_t(M_{12})&=& \Delta_0(M_{12})+\frac{\theta}{2}(P_1\otimes
P_3-P_3\otimes P_1),\cr
\Delta_t(M_{13})&=& \Delta_0(M_{13})-\frac{\theta}{2}(P_1\otimes
P_2-P_2\otimes P_1)\,.\eea
One can see from \eqref{twist-resid} that the generators of the
stability group of $\theta_{\mu\nu}$, i.e. $M_{01}$ which generates
$O(1,1)$ and $M_{23}$ which generates $SO(2)$, both act through the
primitive coproduct. Just as the preservation of translational
symmetry is apparent from the primitive coproduct of the momentum
generators \eqref{twist p}, the invariance under the Lorentz
subgroup $O(1,1)\times SO(2)$ is indicated in the twisted \P
language by the unchanged coproducts of the corresponding
generators. According to \eqref{twist-mix}, the generators whose
coproducts are deformed are those which mix the commutative
directions with the noncommutative ones.

If we wish to discuss various invariances in the context of twisted
\P symmetry, we have to ensure that they hold under finite
transformations, not only infinitesimal ones. To extend the concept
of finite \P transformations to the twisted case, one has to adopt
the dual language of Hopf algebras: the algebra of functions $F(G)$
on the \P group $G$, as a commutative algebra, is dual to $\cal
U(\cal P)$. The algebra $F(G)$ is generated by the elements
$\mathbf{\Lambda}^{\mu}_{\nu}$ and $\mathbf{a}^\mu$, which are
complex-valued functions, such that when applied to suitable
elements of the \P group, they would return the familiar real-valued
entries of the matrix of finite Lorentz transformations,
$\Lambda^\mu_\nu$, or the real-valued parameters of finite
translations, $a^\mu$. For example, if we consider the action of
elements of $F(G)$ on a Lorentz group element
$e^{i\omega^{\alpha\beta}M_{\alpha\beta}}\in G$ (without summation
over $\alpha$ and $\beta$), we obtain
\bea
\mathbf{\Lambda}^\mu_\nu\left(e^{i\omega^{\alpha\beta}M_{\alpha\beta}}\right)&=&\left(\Lambda_{\alpha\beta}(\omega)\right)^\mu_\nu\,,\label{Lorentz_on_Lorentz}\\
\mathbf{a}^\mu\left(e^{i\omega^{\alpha\beta}M_{\alpha\beta}}\right)&=&0\label{Lorentz_on_transl}\,,
\eea
while the action on translation group elements $e^{ia^\alpha
P_\alpha}$ gives
\bea \mathbf{\Lambda}^\mu_\nu\left(e^{ia^\alpha
P_\alpha}\right)&=&0\,,\label{Lorentz_on_transl}\\
\mathbf{a}^\mu\left(e^{ia^\alpha
P_\alpha}\right)&=&a^\mu\,.\label{transl_on_transl}\eea

The duality is preserved after twisting, but with deformed
multiplication in the dual algebra \footnote{A basic property of the
duality is that the coproduct and multiplication of the deformed
Hopf algebra directly influence the multiplication and coproduct,
respectively, of the deformed dual Hopf algebra (see, e.g., Refs.
\cite{ChPr,Maj,ChD}).}. The deformed coproduct of the twisted \P
algebra ${\cal U}_t(\cal P)$ turns into noncommutativity of
translation parameters in the dual $F_\theta (G)$ \cite{oeckl,
Gonera,Kulish}:
\bea\label{finite translations}[\mathbf
{a}^\mu,\mathbf{a}^\nu]&=&i\theta^{\mu\nu}-i\mathbf{\Lambda}^\mu_\alpha\mathbf{\Lambda}^\nu_\beta\theta^{\alpha\beta}\,,\cr
\ \  [\mathbf{\Lambda}^\mu_\nu,
\mathbf{a}^\alpha]&=&[\mathbf{\Lambda}^\mu_\alpha,\mathbf{\Lambda}^\nu
_\beta]=0,\ \ \ \mathbf{\Lambda}^\mu_\alpha, \mathbf{a}^\mu \in
F_\theta (G)\,.\eea
The "coordinates" $x^\mu$, generating the algebra of functions with
$\star$-product ${\cal C}_\theta$, transform by the coaction of the
quantum matrix group (see, e.g., Ref. \cite{ChD}, p. 61):
\be\delta:{\cal C}_\theta\to F_\theta(G)\otimes {\cal C}_\theta\ee
as
\be(x')^\mu=\delta(x^\mu)=\mathbf{\Lambda}^\mu_{\alpha}\otimes
x^\alpha+\mathbf{a}^\mu\otimes 1\,.\ee
The role of the deformed multiplication of "translation parameters"
is to preserve the commutation relation of "coordinates" of the
quantum space,
\be\label{cr-transf}[x'_\mu,x'_\nu]=i\theta_{\mu\nu}\,,\ee
the products being of course taken with the appropriate
multiplication in $F_\theta(G)\otimes {\cal C}_\theta$.

Due to the nontrivial commutation relations \eqref{finite
translations}, in the twisted case the functions $\mathbf{a}^\mu$
are no more complex-valued (though $\mathbf{\Lambda}^\mu_\nu$ still
are, and satisfy \eqref{Lorentz_on_Lorentz} and
\eqref{Lorentz_on_transl}). However, there are elements of the \P
group for which the values of the functions $\mathbf{a}^\mu$ are
still commutative. Such simple cases are the translations and the
trivial Lorentz transformations, $\Lambda^\mu_\nu=\delta^\mu_\nu$,
but they are not the only ones. For definiteness, let us consider
various relevant finite twisted Lorentz transformations, as follows
(again, the $\theta$-matrix is assumed as in \eqref{theta}):

{\it i)} A boost in the commutative direction $x^1$:
\begin{eqnarray}\label{boost comm}
\Lambda_{01}=
\mathbf{\Lambda}\left(e^{\omega^{01}(\alpha)M_{01}}\right)=\left(
\begin{array}{cccc}
\cosh \alpha &\sinh\alpha & 0  & 0 \\
\sinh \alpha & \cosh\alpha & 0  & 0 \\
0 & 0  &1 & 0 \\
0 & 0  & 0 & 1
\end{array}
\right)\,;
\end{eqnarray}

{\it ii)} A rotation between the noncommutative coordinates, $x^2$
and $x^3$:
\begin{eqnarray}\label{rot comm-NC}
\Lambda_{23}=\mathbf{\Lambda}\left(e^{\omega^{23}(\gamma)M_{23}}\right)=\left(
\begin{array}{cccc}
1 &0 & 0  & 0 \\
0 & 1 & 0  & 0 \\
0 & 0  &\cos\gamma & \sin\gamma \\
0 & 0  & -\sin\gamma  &\cos\gamma
\end{array}
\right)\,;
\end{eqnarray}

{\it iii)} A rotation between a commutative and a noncommutative
coordinate, $x^1$ and $x^2$:
\begin{eqnarray}\label{rot comm-NC}
\Lambda_{12}=\mathbf{\Lambda}\left(e^{\omega^{12}(\beta)M_{12}}\right)=\left(
\begin{array}{cccc}
1 &0 & 0  & 0 \\
0 & \cos\beta & \sin\beta  & 0 \\
0 & -\sin\beta  &\cos\beta & 0 \\
0 & 0  & 0 & 1
\end{array}
\right).
\end{eqnarray}

Using the general commutation relations (\ref{finite translations})
applied to the corresponding elements of the Lorentz group, we
obtain in the cases {\it i)} and {\it ii)}
\be\label{comm
transl}[\mathbf{a}^\mu(g),\mathbf{a}^\nu(g)]=[a^\mu,a^\nu]=0\,,\ee
where $g$ is either $e^{\omega^{01}(\alpha)M_{01}}$ or
$e^{\omega^{23}(\alpha)M_{23}}$, respectively. (This result holds
also when time is noncommutative and the $\theta$-matrix contains a
nontrivial block in the upper left corner,
$\theta_{01}=-\theta_{10}=\theta'$.)

In the case {\it iii)}, when commutative and noncommutative
coordinates mix, we obtain
\bea\label{NC
transl}\left[\mathbf{a}^2\left(e^{\omega^{12}(\beta)M_{12}}\right),\mathbf{a}^3\left(e^{\omega^{12}(\beta)M_{12}}\right)\right]&=&[a^2,a^3]=i\theta(1-\cos\beta)\,,\cr
\left[\mathbf{a}^1\left(e^{\omega^{12}(\beta)M_{12}}\right),\mathbf{a}^3\left(e^{\omega^{12}(\beta)M_{12}}\right)\right]&=&[a^1,a^3]=-i\theta\sin\beta\,,\eea
all the other commutators being zero. One can check that for all
other Lorentz transformations mixing commutative and noncommutative
directions, nontrivial commutators of "translation parameters"
arise.

Once more it appears that in the twisted \P context, the Lorentz
transformations corresponding to the stability group of
$\theta_{\mu\nu}$ behave just as in the commutative case, while the
Lorentz transformations mixing the commutative and noncommutative
directions require peculiar noncommuting translations. Remark that
we imposed the Lorentz transformation {\it iii)}, and we ended up
with accompanying noncommuting translations showing up as the {\it
internal mechanism} by which the twisted \P symmetry keeps the
commutator \eqref{cr-transf} invariant\footnote{We can view these
translations as taking upon themselves the noncommutativity which
would be naturally bestowed on the combination of commutative and
noncommutative coordinates. For example, in our case, performing the
Lorentz transformation \eqref{rot comm-NC} in the $x^1,x^2$-plane
would at first sight seem to make both coordinates
$x'^1=x^1\cos\beta+x^2\sin\beta$ and
$x'^2=-x^1\sin\beta+x^2\cos\beta$ noncommutative. With the already
noncommuting $x'^3=x^3$, this would have given three noncommuting
directions in the new system of reference, and two nontrivial
commutators, $[x'^2,x'^3]$ and $[x'^3,x'^1]$. However, the twisted
\P symmetry enforces the appearance of the noncommuting translations
\eqref{NC transl}, which reduce the number of nontrivial commutators
back to one, $[x'^2,x'^3]=i\theta$ (as in \eqref{cr-transf}).}.
While one can still conceive an abstract geometrical meaning for the
transformed generators $x'^\mu \in F_\theta (G)\otimes {\cal
C}_\theta$, it is a conceptual challenge to confer them a physical
meaning.

\subsection{Fields in noncommutative space-time}\label{NC field}

Equipped with these results, let us return to the concept of a
noncommutative field and the action of twisted \P transformations on
it. It was proposed in \cite{CKTZZ} that the construction of
noncommutative fields should be started (as in commutative theories)
from first principles, i.e. the general theory of induced
representations (see, e.g., Ref. \cite{Barut}). In the commutative
case, a classical field is a section of a vector bundle induced by
some representation of the Lorentz group. The natural generalization
of this construction is not succesful in the noncommutative case,
mainly because the universal enveloping algebra of the Lorentz Lie
algebra is not a Hopf subalgebra of the twisted \P algebra. As a
result, Minkowski space $\R^{1,3}$, which in the commutative setting
is realized as the quotient of the \P group by the Lorentz group,
$G/L$ (in an obvious notation), has no noncommutative analogue. For
all mathematical details and the subtle points of the comparison
between the commutative and noncommutative cases, we refer the
reader to Ref. \cite{CKTZZ}. In the same paper a way out was
proposed, which retains Minkowski space but uses finite dimensional
$\cal U(\cal P)$-modules with trivial action of all momentum
generators $P_\mu$ instead of finite dimensional Lorentz-modules.

While entirely agreeing with the analysis of Ref. \cite{CKTZZ}, we
would like to propose here still another interpretation of the
noncommutative field, which is closer to the implications of the
dynamical calculations. Eqs. \eqref{finite translations}, and in
particular \eqref{comm transl} and \eqref{NC transl}, show how
destructive the Lorentz transformations mixing commutative and
noncommutative directions are for the coordinates: the coordinates
become objects belonging to $F_\theta(G)\otimes {\cal C}_\theta$, to
which one can not assign any numbers. The Minkowski space in the
noncommutative setting appears not to have the same deep meaning to
which we are used in Special Relativity, because the commutative and
noncommutative coordinates have distinct properties. Our proposal
is, therefore, to give up the Minkowski space $\R^{1,3}$ in favour
of $\R^{1,1}\times \R^{2}$, but to retain the finite dimensional
Lorentz-modules in the constructions of noncommutative fields.

Specifically, a commutative relativistic field has to carry a
representation of the Lorentz group and at the same time to be a
function of the space-time coordinates $x^\mu \in \R^{1,3}$
\footnote{This statement and the argumentation below it are
presented in an intuitive manner, disregarding mathematical rigour.
For a rigorous treatment we refer the reader to Ref. \cite{CKTZZ}.}.
The consistent construction, such that the actions of the \P group
can be defined on the field, is achieved by the method of induced
representations. The commutative field turns out to be an element of
$C^\infty(\R^{1,3})\otimes V$, where $C^\infty(\R^{1,3})$ is the set
of smooth functions on Minkowski space and $V$ is a Lorentz-module
(a space of representations, bearing the actions of the Lorentz
group). Since the field is defined as a tensor product, the action
of the Lorentz group on it has to go through the coproduct, which in
the commutative case is the primitive coproduct \eqref{primitive}
and this is readily achieved since both $C^\infty(\R^{1,3})$ and $V$
admit actions of the Lorentz group.

In the case of twisted \P algebra, when trying to act with a Lorentz
generator on an element of $C^\infty(\R^{1,3})\otimes V$,
\be \Phi=\sum_i f_i\otimes v_i\,, \ \ \ f_i \in
C^\infty(\R^{1,3})\,,\ \ \ v_i \in V\,,\ee
one has to use the twisted coproduct and at this point the procedure
fails. The twisted coproduct of Lorentz generators \eqref{twist-mix}
contains terms which require the action of the {\it momentum}
operator on the elements of $V$, but $V$ - a Lorentz-module - does
not admit the action of $P_\mu$. This is why it was proposed in
\cite{CKTZZ} to replace the Lorentz-module by a $\cal
U(\cal{P})$-module with trivial actions of the momentum generators.
The consequences of this construction are found in \cite{CKTZZ}.

We propose as a simpler solution to retain $V$ as a Lorentz-module,
but to simply discard the action of all those Lorentz generators
which are not allowed because of the additional terms containing the
inadmissible momentum generator $P_\mu$. Recall from
\eqref{twist-resid} and \eqref{twist-mix} that the generators of the
stability group of $\theta_{\mu\nu}$ still act via the primitive
coproduct, therefore their action on elements of $V$ is not
prevented in any way. Their algebra also closes (it is the Abelian
algebra $o(1,1)\times o(2)$).

To conclude, we propose that the {\it noncommutative field} be in
$C^\infty(\R^{1,1}\times \R^2)\otimes V$, thus {\it carrying
representations of the full Lorentz group, but admitting only the
action of the generators of the stability group of
$\theta_{\mu\nu}$, i.e. $O(1,1)\times SO(2)$}\footnote{Loosely
stated, the difference between the approach of Ref. \cite{CKTZZ} and
the present one is the following: while in Ref. \cite{CKTZZ} the
noncommutative fields were induced by a part of the representations
of the Lorentz group, but carrying the action of all the generators
of the \P algebra through the twisted coproduct, in this paper we
advance the idea of having the noncommutative fields induced by all
the representations of the Lorentz group, but carrying only the
action of the generators of the stability group of
$\theta_{\mu\nu}$. An advantage of the latter approach is that the
finite transformations of the noncommutative fields are readily
obtained.}.

The generalization of this statement to the quantum case is
straightforward: the field $\hat\Phi=\sum_i \hat f_i\otimes v_i$
becomes an operator through $\hat f_i=\hat A_i\otimes g_i$ which
belong to ${\cal O}\otimes C^\infty(\R^{1,1}\times \R^2)$, where
$\cal O$ is an algebra of field operators acting on the Hilbert
space of states. The product of the field operators is not
influenced by the twist, while the functions of
$C^\infty(\R^{1,1}\times \R^2)$ are multiplied by the
$\star$-product:
\be\label{quantum-field}(\hat A\otimes g)(\hat B\otimes h)= \hat
A\hat B\otimes g\star h\,,\ \ \ \hat A,\hat B\in {\cal O}\,,\ \ \
g,h\in C^\infty(\R^{1,1}\times \R^2)\,.\ee
 The Lorentz-module $V$ is in no way affected by the
quantization. What is different compared with the commutative case
is that now the field picks up its $x$-dependence from
$\R^{1,1}\times \R^2$ instead of the Minkowski space, which is in
full agreement with the dynamical calculations. Again, only the
action of the generators of the stability group of $\theta_{\mu\nu}$
is allowed and it goes through the primitive coproduct. Since the
quantum field $\hat\Phi$ carries a representation of the Lorentz
group through $v_i$, the field operators $\hat A_i$ will carry in
their turn corresponding Lorentz representation indices. This,
together with the usual product in the algebra of operators $\cal
O$, make the Hilbert space of states (in essence, the Fock space)
identical to the one of the commutative QFT\footnote{Consequently,
the spin-statistics relation holds just as in the commutative case.
}.

\section{Conclusions}

In this paper we have studied the confrontation of the Lorentz
symmetry, the residual $O(1,1)\times SO(2)$ symmetry and the twisted
\P symmetry in  noncommutative QFT with constant antisymmetric
parameter $\theta_{\mu\nu}$. Based on Pauli's Theorem
\cite{Pauli,KN-pauli}, we concluded that the Lorentz group can not
provide a symmetry for NC QFT. We have presented a new dynamical
result, the Tomonaga-Schwinger equation in the interaction picture
of NC QFT, which supports the previous computations in various
models, showing the infinite nonlocality in the noncommutative
directions, the emergence of the light-wedge causality condition and
the symmetry of NC QFT under the stability group of
$\theta_{\mu\nu}$, $O(1,1)\times SO(2)$. This result is general and
of significance for building up concrete noncommutative models.

Persuaded that the dynamical calculations and the symmetry arguments
have to match each other in NC QFT as in any other physical theory,
we embarked upon deepening our understanding of what is meant by
twisted \P invariance. Following the proposal of Ref. \cite{CKTZZ}
to approach the definition of the noncommutative fields starting
from the method of induced representations, we proposed in Section
4.2 a new interpretation for the noncommutative fields. With this
construction, the meaning of the twisted \P symmetry in NC QFT
becomes transparent: it represents actually the invariance with
respect to the stability group of $\theta_{\mu\nu}$, while the
quantum fields still carry representations of the full Lorentz group
and the Hilbert space of states has the richness of particle
representations of the commutative QFT.

Thus, the twisted \P symmetry and the invariance under the stability
group of $\theta_{\mu\nu}$ peacefully coexist in NC QFT. Lorentz
symmetry can not be achieved  with constant noncommutativity
parameter, therefore noncommutative QFT can not be interpreted as
indistinguishable from commutative QFT.

\vskip 0.3cm {\bf{Acknowledgements}}

We are much grateful to Peter Pre\v{s}najder, Shahin Sheikh-Jabbari
and Ruibin Zhang for useful discussions. A.T. acknowledges the
project no. 121720 of the Academy of Finland.

\vskip1cm

{\Large \bf Appendix} \vskip1cm
\appendix

\section{Light-wedge configuration}

Instead of showing that, in general,
\be \left[ \mathcal H_{int}(x)_\star,\mathcal H_{int}(y)_\star
\right] \ee
does not vanish identically for $(x^0-y^0)^2<( \vect x-\vect y)^2$
we choose a simpler commutator
$$\bigl[\phi(x),\phi(y)\star\phi(y)\bigr].$$ We know that it vanishes for $(x^0-y^0)^2<(  x^1- y^1)^2$, but
we now want to show that it does not necessarily vanish
for$(x^0-y^0)^2<( \vect x-\vect y)^2$.

For this purpose we consider it in the form
\begin{align}\label{commutator}
\bigl[\phi(x),\phi(y)\star\phi(y)\bigr] = i
\Delta(x-y)\star\phi(y)+i\phi(y)\star\Delta(x-y) \,.
\end{align}
Writing the field in terms of the Fourier transform
\be\phi(y)= \sum_l e^{il\cdot y}\phi(l)\,,\ee
\eqref{commutator} is proportional to
\begin{align}\label{Deltas}
\sum_l e^{il\cdot y}\bigl[ &\Delta(x^0-y^0,x^1-y^1,x^2-y^2+\theta l_3,x^3-y^3-\theta l_2)+ \\
&\Delta(x^0-y^0,x^1-y^1,x^2-y^2-\theta l_3,x^3-y^3+\theta l_2)\bigl]
\,.
\end{align}
For this to vanish the quantity in brackets should vanish for every
$l$ since all $\phi(l)$ are linearly independent. We will now show
that this does not happen in a special configuration, where $x^1-y^1
=x^3-y^3=0$ and $l_2 = 0$, which implies the light cone condition
$(x^0-y^0)^2<(  x^2- y^2)^2$.

Let us choose $x^2-y^2+\theta l_3 = 0$. Now in the first term of
\eqref{Deltas} all the space coordinates vanish and
$-(x^0-y^0)^2<0$, i.e. the vector is timelike and this term survives
in the integration. Therefore the commutator
$\bigl[\phi(x),\phi(y)\star\phi(y)\bigr]$ does not vanish for
$(x^0-y^0)^2<( \vect x-\vect y)^2$ even though it vanishes for
$$(x^0-y^0)^2<(  x^1- y^1)^2$$ and we can conclude that the
commutator does not vanish for a space-like separation if the
light-wedge condition is not met.

\end{document}